\documentclass[preprint,eqsecnum,preprintnumbers,nofootinbib,byrevtex,prd,aps,showpacs,showkeys,groupedaddress,
floatfix]{revtex4}
\usepackage{bm}
\usepackage{graphics}
\usepackage{graphicx}
\usepackage{epsfig}
\usepackage{amssymb}
\usepackage{amsmath}

\newcommand\nn{\nonumber}
\newcommand\ba{\begin{eqnarray}}
\newcommand\ea{\end{eqnarray}}

\begin{document}
\title{Bound State  Solution of the  Klein-Fock-Gordon
equation  with the Hulth\'en  plus a Ring-Shaped like potential
within SUSY quantum mechanics}
\author{A.~I.~Ahmadov$^{1,2}$~\footnote{E-mail: ahmadovazar@yahoo.com}}
\author{Sh.M.~Nagiyev$^{3}$~\footnote{E-mail: shakir.m.nagiyev@gmail.com}}
\author{M.V.~Qocayeva$^{3}$~\footnote{E-mail: mefkureqocayeva@yahoo.com}}
\author{K.~Uzun$^{4}$~\footnote{E-mail: $\mbox{kubrakaraoglu}_-\mbox{2561@hotmail.com}$}}
\author{V.A.~Tarverdiyeva$^{3}$~\footnote{E-mail:vefa.tarverdiyeva@mail.ru}}
\affiliation{$^{1}$ Department of Theoretical  Physics, Baku State
University,\\ Z. Khalilov st. 23, AZ-1148, Baku, Azerbaijan}
\affiliation{$^{2}$ Institute for Physical Problems, Baku State
University,\\ Z. Khalilov st. 23, AZ-1148, Baku, Azerbaijan}
\affiliation{$^{3}$ Institute of Physics, Azerbaijan National
Academy of Sciences,\\ H. Javid Avenue,131, AZ-1143, Baku,
Azerbaijan} \affiliation{$^{4}$\ Department of Physics, Karadeniz
Technical University, 61080, Trabson, Turkey}

\date{}
\begin{abstract}

In this paper,  the bound state solution of the modified
Klein-Fock-Gordon equation is obtained for the Hulth\'en plus
ring-shaped lake potential by using the developed scheme to overcome
the centrifugal part. The energy eigenvalues and corresponding
radial and azimuthal  wave functions  are defined for any $l\neq0$
angular momentum case on the conditions that scalar potential is
whether equal and nonequal to vector potential, the bound state
solutions of the Klein-Fock-Gordon equation of the Hulth\'en plus
ring-shaped like potential are obtained by Nikiforov-Uvarov (NU) and
supersymmetric quantum mechanics (SUSYQM)  methods. The equivalent
expressions are obtained for the energy eigenvalues, and the
expression of radial wave functions transformations to each other is
revealed owing to both methods. The energy levels and the
corresponding normalized eigenfunctions are represented in terms of
the Jacobi polynomials for arbitrary $l$ states. A closed form of
the normalization constant of the wave functions is also found. It
is shown that the energy eigenvalues and eigenfunctions are
sensitive to $n_r$ radial and $l$ orbital quantum numbers.
\end{abstract}

 \maketitle

\section{Introduction}

Since the early years of quantum mechanics (QM), the study of
exactly solvable problems for some special potentials has aroused
considerable interest in theoretical physics. In addition, since the
wave function contains all necessary knowledge for the full
description of a quantum system, so an analytical solution of the
wave equations is of quite high significance in quantum mechanics
~\cite{Greiner, Bagrov}.

Since the exact solutions of the Klein-Fock-Gordon (KFG) equation
with any potential play an important role in relativistic quantum
mechanics \cite{Greiner,Bagrov}, there are many discussions about
the KFG equation with physical potentials by using different
methods. KFG equation is the well known relyativistic wave equation
that describes spin zero particles, as psevdoscalar pions. For
example, the s-wave KFG equation with the vector Hulth\'en-type
potential was treated by standard method \cite{Znojil}, the same
problem but with both vector and scalar Hulth\'en-type potentials
was later discussed in \cite{Adame,Chen}, the scattering state
solutions of the s-wave KFG equation with vector and scalar
Hulth\'en potentials are obtained for regular and irregular boundary
conditions in \cite{Talukar}. Chetouani $et$ $al.$ successfully
solved the Green function for the KFG operator with these two
potentials by using the path-integral approach \cite{Chetouani}.

Many methods were developed and has been used successfully in
solving the non-relativistic and relativistic   wave equations in
the presence of some well known potentials. Such as supersymmetry
(SUSY)~\cite{Cooper1,Cooper2,Morales}, factorization~\cite{Dong1} ,
Laplace transform approach~\cite{Arda} and the path integral
method~\cite{Cai}, shifted 1/N expansion approach ~\cite{Tang,Roy}
for solving radial and azimuthal part of the wave equations exactly
or quasi-exactly for $l\neq0$ within different potentials. An other
method known as the Nikiforov-Uvarov (NU) method ~\cite{Nikiforov}
was proposed for solving the wave equations analytically.

In works
\cite{Chen,Oluwadre,Cheng,Mehmet,Sever,Yuan,Qiang,Dong2,Dong3,Saad,Boztosun},
the scalar potential is equal and non-equal to the vector potential
have been assumed to obtain the bound states of the KFG equation
with some typical potential by using the ordinary quantum mechanics.
It is very significant to notice that KFG equation for the
Ring-Shaped potential is fully studied in Ref.\cite{Dong2}

In order to give correction for non-relativistic quantum mechanics,
the investigation of relativistic wave equations, which is invariant
under Lorentz transformation, is required by the description of
phenomena at high energies.\cite{Bagrov}.

If we consider the case where the interaction potential is not
enough to create particle-antiparticle pairs, we can apply the KFG
equation to the treatment of a zero-spin particle and apply the
Dirac equation to that of a $1/2$-spin particle. When particle is in
a strong field, then will interesting  to consider the relativistic
equations, so we can if possible extract the correction to
non-relativistic quantum mechanics. Since it has been extensively
used to describe the bound and continuum states of the interacting
systems, it would be quite curious and significant investigation to
the relativistic bound states of the arbitrary $l$-wave KFG equation
with Hulth\'en potential plus a ring-shaped like potential.

The Hulth\'en potential is one of the important short-range
potentials in physics, extensively using to describe the bound and
continuum states of the interaction systems. It has been applied to
the several research areas such as nuclear and particle physics,
atomic physics, condensed matter and chemical physics, so the
analyzing relativistic effects for a particle under this potential
could become significant, especially for strong coupling. Therefore
this problem has attracted a great deal of interests in solving the
KFG equation with the Hulth\'en potential.

The  Hulth\'en potential is defined by~\cite{Hulten1,Hulten2}
\ba
V(r)=-\frac{Ze^2\delta e^{-\delta r}}{(1-e^{-\delta r})}
\label{a1}
\ea

At small values of the radial coordinate $r$, the Hulth\'en
potential behaves like a Coulomb potential,  whereas for large
values of $r$ it decreases exponentially so that its influence for
bound state is smaller than, that of Coulomb potential. In contrast
to the Hulth\'en potential, the Coulomb potential  is analytically
solvable for any $l$ angular momentum. Take into account of this
point will be very interesting and important solving KFG equation
for the Hulth\'en plus ring-shaped like potential for any $l$ states
within ordinary  and supersymmetric quantum mechanics.

Unfortunately, for an arbitrary $l$-states ($l\neq 0$), the KFG
equation does not get an exact solution due to the centrifugal term.
But many research are show the power and simplicity of NU method in
solving central and noncentral potentials~\cite{Badalov1,
Badalov2,Badalov3,Ahmadov1, Ahmadov2, Ahmadov3, Badalov4, Ahmadov4,
Ahmadov5} for arbitrary $l$ states. This method is based on solving
the second-order linear differential equation by reducing to a
generalized equation of hypergeometric-type which is a second-order
type  homogeneous differential equation with polynomials
coefficients of degree not exceeding the corresponding order of
differentiation.

It should be noted that the nature of the radial function at the
origin, especially for singular potentials was comprehensively
studied by Khelashvili et al.
~\cite{Khelashvili1,Khelashvili2,Khelashvili3,Khelashvili4,Khelashvili5,Khelashvili6}.
While the Laplace operator is defined in spherical coordinates, the
exact derivation of the radial wave  equation displays the
appearance of a delta function term. As a result, regardless of the
behavior potential, the additional constraint is imposed on radial
wave function in the form of a vanishing boundary condition at the
origin.

The combined  potential  considering  in this study is obtained by
adding Hulth\'en potential term to Ring-Shaped potential as:

\ba
V(r,\theta)=\frac{\hbar^2}{2M}\left[-\frac{2M}{\hbar^2}\frac{Ze^2\delta
e^{-\delta r}}{(1-e^{-\delta
r})}+\frac{\beta'}{r^2sin^2\theta}+\frac{\beta
cos\theta}{r^2sin^2\theta}\right]. \label{a2} \ea

The non-central potentials are needed to obtain better results than
central potentials about the dynamical properties of the molecular
structures and interactions. Ring-shaped potentials can be used in
quantum chemistry to describe the ring shaped organic molecules such
as benzene and in nuclear physics to investigate the interaction
between deformed pair of nucleus and spin orbit coupling for the
motion of the particle in the potential fields.

This potential also is used as a mathematical model in the
description of diatomic molecular vibrations and it constitutes a
convenient model for other physical situations.

Therefore, it would be interesting and important to solve the
relativistic radial  and azimuthal KFG  equation  for Hulth\'en plus
ring-shaped like potential for $l\neq0$, since it has been
extensively used to describe the bound and continuum states of the
interacting systems.

Thus, the main purpose of our  investigation is the analytical
solution of modified KFG  equation for the Hulth\'en plus
ring-shaped potential within ordinary  quantum mechanics using
Nikiforov-Uvarov (NU) method ~\cite{Nikiforov} and in SUSY quantum
mechanics the shape invariance concept that was introduced by
Gendenshtein ~\cite{Gendenshtein1,Gendenshtein2} by using a novel
improved scheme to overcome centrifugal term and found the energy
eigenvalues and corresponding radial and azimuthal wave functions
for any $l$ orbital angular momentum case.

The rest of the present work is organized as follows. Bound-state
solution of the radial KFG  equation for Hulth\'en potential by NU
method within ordinary quantum mechanics is provided in Section
\ref{br}. In Section \ref{ar}, we present the solution of
angle-dependent part of the KFG equation. In Section \ref{sr} we
present the solution of KFG equation for Hulth\'en potential within
SUSY quantum mechanics and the numerical results for energy levels
and the corresponding normalized eigenfunctions are presented in
Section \ref{nr}. Finally, some concluding remarks are stated in
Section \ref{cr}.

\section{BOUND STATE SOLUTION OF THE RADIAL KLEIN-FOCK-GORDON EQUATION}\label{br}

Two various type potentials can be introduced into this equation
because KFG equation contains two objects; the four-vector linear
momentum operator and the scalar rest mass. The first one is a
vector potential (V), introduced via minimal coupling and the second
one is a scalar potential (S) introduced via scalar
coupling\cite{Greiner}. Hence, they allow one to introduce two types
of potential coupling which are the four vector potential (V) and
the space-time scalar potential (S).

In spherical coordinates, the KFG equation with  scalar potential
$S(r,\theta)$ and vector potential $V(r,\theta)$ can be written in
the following form in natural units ($\hbar=c=1$)
\ba
  [-\nabla ^2 +(M+S(r,\theta))^2]\psi (r,\theta,\phi)=[E-V(r,\theta)]^2\psi
  (r,\theta,\phi),
\label{a3}
\ea
where $E$ is the relativistic energy of the system and $M$ denotes
the rest mass of a scalar particle.

For separation of radial and angular parts of the wave function for
the stationary KFG equation with Hulth\'en plus ring-shaped
potential we use following wave function \ba \psi
(r,\theta,\phi)=\frac {\chi (r)}{r}\Theta
(\theta)e^{im\phi},~~m=0,\pm 1,\pm 2,\pm 3 ... \label{a4} \ea

and substituting this into Eq.\eqref{a3} leads to the following
second-order differential equations \ba
\chi^{''}(r)+\left[(E^2-M^2)-2(MS(r)+EV(r))+(V^{2}(r)-S^{2}(r))-\frac{l(l+1)}{r^2}\right]\chi(r)=0, \nn \\
\label{a5} \ea
\ba \Theta ^{''}(\theta )+cot\theta ~\Theta ^{'}(\theta)
+\left[\frac {-2}{sin^2\theta }\left({(M+E)}(\beta^{'}+\beta
cos\theta)+m^2\right) +\lambda \right]\Theta(\theta) =0. \label{a6}
\ea \noindent It should be noted that in Eq.\eqref{a6}, the scalar
ring-shaped potential is taken to equal with vector potential
$V(r,\theta)=S(r,\theta)$.

\noindent If  we take vector and scalar potentials as the general
Hulth\'en potential in this form

\ba V(r)= -\frac{V_0e^{-\delta r}}{(1-e^{-\delta r})},\,\,\,\,\,\,
S(r)= -\frac{S_0e^{-\delta r}}{(1-e^{-\delta r})}, \label{a7} \ea
then Eq.\eqref{a5} becomes \ba
\chi^{''}(r)+\left[(E^2-M^2)+2\left(\frac{(MS_0+EV_0)e^{-\delta
r}}{1-e^{-\delta
r}}\right)+2\left(\frac{(V^{2}_0-S_0^{2})e^{-2\delta
r}}{(1-e^{-\delta r})^2}\right)-\frac{l(l+1)}{r^2}\right]\chi(r)=0, \nn \\
\label{a8} \ea

The effective Hulth\'en potential is defined in this form:

\ba V_{\rm eff}(r)=-2\left(\frac{(MS_0+EV_0)e^{-\delta
r}}{1-e^{-\delta
r}}\right)+2\left(\frac{(S^{2}_0-V_0^{2})e^{-2\delta
r}}{(1-e^{-\delta r})^2}\right)+\frac{l(l+1)}{r^2}\label{a118} \ea

It is known that for this potential the KFG equation can be solved
exactly using suitable approximation scheme to deal with the
centrifugal term.

Therefore, in this research study, we attempt to use the following
improved approximation scheme to deal with the centrifugal term. In
order to solve Eq.\eqref{a8} for $l\neq0,$ we should make an
approximation for the centrifugal term. When $\delta r < < 1$, we
use an improved approximation scheme~\cite{Wen1,Wei,Dong6} to deal
with the centrifugal term,
\ba
\frac {1}{r^2}\approx {\delta^2}\left[C_0+\frac{e^{-\delta
r}}{(1-e^{-\delta r})^2} \right],
\label{a11}
\ea
\noindent where the parameter $C_0=\frac{1}{12}$ (Ref.~\cite{Jia1})
is a dimensionless constant. However, when $C_0=0$, the
approximation scheme becomes the convectional approximation scheme
suggested by Greene and Aldrich~\cite{Greene}.

Now for applying  NU method, Eq.\eqref{a8} should be rewritten as
the hypergeometric type equation form presenting below: \ba
\chi^{''}(s)+\frac{\tilde{\tau}}{\sigma}\chi^{'}(s)+\frac{\tilde{\sigma}}{\sigma^2}\chi(s)=0,
\label{a14} \ea
The Eq.\eqref{a8} can be further simplified using a new variable
$s=e^{-\delta
 r}$. Taking into account, that here $r \in[0,\infty)$ and $s\in[1,0]$, then we obtain:

\begin{eqnarray}
\chi''(s)+\chi
'(s)\frac{1}{s}+\biggl[\frac{-\varepsilon^2}{s^2}+\frac{\alpha^2}{s(1-s)}-\frac{\beta^2}{(1-s)^2}-\frac{l(l+1)}{s^2\delta^2
r^2}\biggr]\chi(s)=0, \label{a9}
\end{eqnarray}

\noindent where we use the following notations for bound states
\ba \varepsilon=\frac{\sqrt{M^2-E^2}}{\delta}~,~~\alpha
=\frac{\sqrt{2EV_0+2MS_0}}{\delta}~,~~~\beta
=\frac{\sqrt{S_0^2-V_0^2}}{\delta}. \label{a10} \ea
For the bound states, should $E<M$, $\varepsilon >0$. The boundary
conditions for Eq.\eqref{a5}  are $\chi(0)=0$ and $\chi(\infty)=0$.
Having inserting  Eq.\eqref{a11} in Eq.\eqref{a9} and  after such
manipulations we obtain:

\ba \chi''(s)+\chi
'(s)\frac{1-s}{s(1-s)}+\biggl[\frac{1}{s(1-s)}\biggr]^2\biggl[-\varepsilon^2(1-s)^2+
\alpha^2 s(1-s)-\beta^2 s^2 \nn \\
-l(l+1)C_0(1-s)^2-l(l+1)s\biggr]\chi(s)=0, \label{a15} \ea

Now, NU method can be successfully applied to define the eigenvalues
of energy. By comparing Eq.\eqref{a15} with Eq.\eqref{a14} we can
define the followings
\ba \tilde{\tau} (s) = 1 - s,\sigma (s) = s(1 - s),\\
\tilde{ \sigma} (s) =-\varepsilon^2(1-s)^2 +\alpha^2s(1-s)-\beta^2
s^2-\lambda C_0(1-s)^2-\lambda s. \label{a17} \ea

\noindent If we take the following factorization
\ba \chi (s)=\phi (s)y(s), \label{a18} \ea for the appropriate
function $\phi (s)$ the Eq.\eqref{a15}  takes the form of the well
known hypergeometric-type equation,
\ba
\sigma (s) y^{''} (s) + \tau (s) y^{'} (s) +\bar{\lambda} y(s)=0.
\label{a19}
\ea
The appropriate $\phi (s)$ function must satisfy the following
condition
\ba
\frac { \phi ^{'} (s)}{\phi (s)}=\frac {\pi (s)}{\sigma (s)},
\label{a20}
\ea
\noindent where $\pi (s)$, the polynomial of degree at most one, is
defined as
\ba \pi(s)= \frac{{ \sigma' -\tilde{\tau }}}{2} \pm \sqrt
{\left(\frac{{ \sigma' -\tilde{\tau} }}{2}\right)^2 -\tilde{\sigma}
+k\sigma}. \label{a21} \ea
Finally the equation, where $y(s)$ is one of its solutions, takes
the form known as hypergeometric-type if the polynomial
$\bar\sigma(s)=\tilde\sigma(s)+\pi^2(s)+\pi(s)[\tilde\tau(s)-\sigma^{'}(s)]+\pi^{'}(s)\sigma(s)$
is divisible by $\sigma(s)$, i.e., $\bar\sigma=\bar \lambda
\sigma(s)$.
The constant $\bar \lambda$ and polynomial $\tau (s)$ in
Eq.\eqref{a19} defined as
\ba \bar{\lambda} =k+\pi^{'}(s) \label{a22} \ea
and
\ba
\tau (s)=\tilde{\tau} (s) +2\pi (s),
\label{a23}
\ea
respectively. For our problem, the $\pi (s)$ function is written as
\ba
\pi(s)= \frac{{ - s}}{2} \pm \sqrt {s^2 [a - k] - s[b -k] + c}
\label{a24}
\ea
where the values of the parameters are
$$
 a = \frac{1}{4} +\varepsilon ^2 + \alpha^2+ \beta^2 + \lambda C_0,
$$
$$
 b = 2\varepsilon ^2 + 2\lambda C_0 +\alpha^2-\lambda,
$$
$$
 c = \varepsilon ^2  + \lambda C_0~.
$$

The constant parameter $k$ can be found complying with the condition
that the discriminant of the expression Eq.\eqref{a24} under the
square root is equal to zero. Hence we obtain
\ba
k_{1,2}  = (b - 2c) \pm 2\sqrt {c^2  + c(a - b)}~.
\label{a25}
\ea
When the individual values of $k$ given in Eq.\eqref{a24} are
substituted into Eq.\eqref{a23}, the four possible forms of $\pi(s)$
are written as follows
\ba
\pi (s) = \frac{{ - s}}{2} \pm \left\{ \begin{array}{l}
(\sqrt c  - \sqrt {c + a - b} )s - \sqrt c \,\,\,for\,\,\, k = (b - 2c) + 2\sqrt {c^2  + c(a - b)} , \\
(\sqrt c  + \sqrt {c + a - b} )s - \sqrt c \,\,\, for\,\,\, k = (b - 2c) - 2\sqrt {c^2  + c(a - b)} . \\
\end{array} \right.
\label{a26}
\ea
The polynomial $\pi(s)$ have four possible form according to NU
method, but we select the one for which the function $\tau (s)$  has
the negative derivative. Another  forms are not suitable physically.
Therefore, the appropriate function $\pi(s)$ and $\tau(s)$ are
\ba
\pi(s)=\sqrt{c}-s\left[\frac{1}{2}+\sqrt{c}+\sqrt{c+a-b}\right],
\label{a27}
\ea
\ba
\tau (s) = 1 +2\sqrt c -2s \left[ 1+\sqrt{c+a-b}\right],
\label{a28}
\ea
for
\ba
k  = (b - 2c) -2\sqrt {c^2  + c(a - b)}.
\label{a29}
\ea
Also by Eq.\eqref{a21} we can define the constant $ \bar{\lambda} $
as
\ba
\bar{ \lambda}=b-2c- 2\sqrt {c^2  + c(a - b)}  - \left[\frac{1}{2} +
{\sqrt{c} + \sqrt {c + a - b} }\right].
\label{a30}
\ea
Given a nonnegative integer $n_r$, the hypergeometric-type equation
has a unique polynomials solution of degree $n_r$ if and only if
\ba
\bar{\lambda}=\bar{\lambda}_{n_r}=-n_{r}\tau'-\frac{n_r(n_r-1)}{2}\sigma
'', (n_r=0,1,2...), \label{a31} \ea
and $\bar{\lambda}_m\neq\bar{\lambda}_n $ for
~$m=0,1,2,...,n_{r}-1$, then it follows that,

$$
\bar{\lambda} _{n_{r}}  = b-2c- 2\sqrt {c^2  + c(a - b)}
-\left[\frac{1}{2} + {\sqrt{c} + \sqrt {c + a - b} }\right]
$$
\ba
= 2n_r\left[ {1 +\left( {\sqrt c  + \sqrt {c + a - b} } \right)}
\right] + n_r(n_r - 1).
\label{a32}
\ea
We can solve Eq.\eqref{a32} explicitly for $c$ by using the relation
 $c=\epsilon ^2  + \lambda C_0$ which brings
\ba \varepsilon^{2}  = \left[
\frac{\alpha^2-\lambda-1/2-n_r(n_r+1)-(2n_r+1)\sqrt{\frac{1}{4}+\beta^2+\lambda}
}{ 2n_r + 1 + 2\sqrt{\frac{1}{4}+\beta^2+\lambda}} \right]^2
-\lambda C_0, \label{a33} \ea
After inserting $\varepsilon ^2$ into Eq.\eqref{a10} for energy
levels in more common  case $V(r)\neq S(r)$ we find
\ba M^2-E_{n_{r},l}^2 = \left[
\frac{\alpha^2-\lambda-1/2-n_r(n_r+1)-(2n_{r}+1)\sqrt{\frac{1}{4}+\beta^2+\lambda}
}{ 2n_r + 1 + 2\sqrt{\frac{1}{4}+\beta^2+\lambda}}\cdot\delta
\right]^2 - \nn \\ l(l+1) C_0 \delta^2. \label{a34} \ea

\noindent In case $V_0=S_0$, then for energy spectrum we obtain:
\ba M^2-E_{n_{r},l}^2 = \left[
\frac{\alpha^2-\lambda-1/2-n_r(n_r+1)-(2n_r+1)\sqrt{\frac{1}{4}+\lambda}
}{2n_r + 1 + 2\sqrt{\frac{1}{4}+\lambda}}\cdot\delta \right]^2
-l(l+1) C_0 \delta^2= \nn \\
\left[\frac{\alpha^2}{2(n_r+l+1)}-\frac{(n_r+l+1)}{2} \right]^2
\cdot\delta^2-l(l+1) C_0 \delta^2. \label{a35} \ea

\noindent In this case $\beta^2 =0$, but $\alpha^2
=\frac{2V_0(E+M)}{\delta^2}$. If we take $C_0=0$ in Eq.\eqref{a34}
and  Eq.\eqref{a35} then we directly obtain results
\cite{Yuan,Qiang}.

\noindent In case $V(r)=-S(r)$, then for energy spectrum we obtain:
\ba M^2-E_{n_{r},l}^2 = \left[
\frac{\alpha'^2-\lambda-1/2-n(n+1)-(2n+1)\sqrt{\frac{1}{4}+\lambda}
}{ 2n_r + 1 + 2\sqrt{\frac{1}{4}+\lambda}}\cdot\delta \right]^2
-l(l+1) C_0 \delta^2= \nn \\
\left[ \frac{\alpha'^2}{2(n+l+1)}-\frac{(n+l+1)}{2}
\right]^2\cdot\delta^2-l(l+1) C_0 \delta^2. \label{a36} \ea

\noindent In this case also $\beta^2 =0$, but $\alpha'^2\neq
\alpha^2$, here

\ba \alpha'^2 =\frac{{2V_0(E-M)}}{\delta^2}\label{a37} \ea

For fully investigation, we also studied   non-relativistic limit of
the  formula for the energy spectrum. When $V(r)=S(r)$, then
Eq.\eqref{a5} reduces to a Schr\"{o}dinger equation for the
potential $2V(r)$. In this case from Eq.\eqref{a5} we directly
obtain result\cite{Ahmadov3}.

The energy levels $ E_{n_r,l}$ is determined by the energy equation
Eqs. (\ref{a34}-\ref{a36}), which is rather complicated
transcendental equation.

Now, applying the NU-method we can obtain the radial eigenfunctions.
After substituting $\pi(s)$ and $\sigma(s) $ into  Eq.\eqref{a20}
and solving first order differential equation, it is easy to obtain
\ba \phi (s)=s^{\sqrt c}(1-s)^K, \label{a38} \ea
where $K=1/2+\sqrt{\frac{1}{4}+\beta^2+l(l+1)}$.

Furthermore, the other part of the wave function $y_n(s)$ is the
hypergeometric-type function whose polynomial solutions are given by
Rodrigues relation
\ba y_{n}(s) = \frac {C_{n}}{\rho (s)} \frac{{d^{n} }}{{ds^{n}
}}\left[ \sigma ^{n}(s)\rho (s) \right], \label{a39} \ea
where $C_n$ is a normalizing constant and $\rho (s)$ is the weight
function which is the solutions of the Pearson differential
equation. The Pearson  differential equation and $\rho(s)$ for this
problem have the form,
\ba (\sigma \rho )^{'} =\tau \rho , \label{a40} \ea
\ba \rho(s) =(1 -s)^{2K - 1} s^{2\sqrt c }, \label{a41} \ea
respectively.

Substituting  Eq.\eqref{a41}  into  Eq.\eqref{a39} then we get
\ba y_{n_{r}}(s) = C_{n_{r}}(1 - s)^{1 - 2K} s^{2\sqrt c }
\frac{{d^{n_{r}} }}{{ds^{n_{r}} }}\left[ {s^{2\sqrt c  + n_{r}} (1 -
s)^{2K - 1 + n_{r}} } \right]. \label{a42} \ea
Then by using the following definition of the Jacobi  polynomials
~\cite{Abramowitz}
\ba P_n^{(a,b)} (s) = \frac{( - 1)^n }{n!2^n (1 - s)^a (1 +
s)^b}\frac{d^n }{ds^n }\left[ {(1 - s)^{a + n} (1 + s)^{b + n} }
\right], \label{a43} \ea
we can write
\ba P_n^{(a,b)} (1-2s) = \frac{C_n}{ s^a (1 - s)^b}\frac{d^n }{ds^n
}\left[s^{a+n}(1-s)^{b+n}\right], \label{a44} \ea
and
\ba \frac{d^n }{ds^n }\left[s^{a+n}(1-s)^{b+n}\right]=C_n s^a (1 -
s)^b P_n^{(a,b)} (1-2s). \label{a45} \ea

If we use the last equality in Eq.\eqref{a42}, we can write
\ba y_{n_{r}}(s) = C_{n_{r}} P_{n_{r}}^{(2\sqrt{c},2K-1)} (s).
\label{a46} \ea
Substituting $\phi (s)$ and $y_{n_{r}}(s)$ into  Eq.\eqref{a20}, we
obtain
\ba \chi _{n_{r}}(s)=C_{n_{r}}s^{\sqrt c}(1-s)^K
P_{n_{r}}^{(2\sqrt{c},2K-1)} (s). \label{a47} \ea
Using the following definition of the Jacobi
polynomials~\cite{Abramowitz}
\ba P_n^{(a,b)} (s) = \frac{{\Gamma (n + a + 1)}}{{n!\Gamma (a +
1)}}\mathop F\limits_{21} \left( { - n,a + b + n + 1,1 + a;s}
\right), \label{a48} \ea
we  are able to write  Eq.\eqref{a47} in terms of hypergeometric
polynomials as
\ba \chi_{n_{r}} (s)=C_{n_{r}}s^{\sqrt c}(1-s)^{K}\frac{\Gamma
(n_{r}+2\sqrt c+1)}{n_{r}!\Gamma (2 \sqrt c+1)} \mathop
F\limits_{21} \left( { - n_{r},2 \sqrt c +2K+n_{r},1 +2 \sqrt
c;s}\right). \label{a49} \ea
The normalization constant $C_{n_{r}}$ can be found from
normalization condition
\ba \int\limits_0^\infty |R(r)|^2r^2dr=\int\limits_0^\infty |\chi
(r)|^2 dr=\frac{1}{\delta}\int\limits_0^1\frac{1}{s}|\chi (s)|^2
ds=1, \label{a50} \ea
by using the following integral formula~\cite{Abramowitz}

$$
\int\limits_0^1 {(1 - z)^{2(\delta  + 1)} z^{2\lambda  - 1} }
\left\{ {\mathop F\limits_{21} ( - n_{r},2(\delta  + \lambda  + 1) +
n_{r},2\lambda  + 1;z)} \right\}^2 dz
$$
\ba
 = \frac{{(n_{r} + \delta  + 1)n_{r}!\Gamma (n_{r} + 2\delta  + 2)\Gamma (2\lambda )\Gamma (2\lambda  + 1)}}{{(n_{r} + \delta  + \lambda  + 1)\Gamma (n_{r} + 2\lambda  + 1)\Gamma (2(\delta  + \lambda  + 1) + n_{r})}},
 \label{a51}
\ea
here $ \delta  > \frac{{ - 3}}{2}$\,\,\, and\,\,\, $\lambda >0 $.
After simple calculations, we obtain normalization constant as
\ba C_{n_r}=\sqrt{\frac{n_{r}!2\sqrt c(n_{r}+K+\sqrt c)\Gamma
(2(K+\sqrt c)+n_{r})}{b(n_{r}+K)\Gamma (n_{r}+2\sqrt c+1)\Gamma
(n_{r}+2K)} }. \label{a52} \ea

\section{\bf Solution of Azimuthal Angle-Dependent Part of the Klein-Fock-Gordon equation}\label{ar}

We may also derive the eigenvalues and eigenvectors of the azimuthal
angle dependent part of the KFG equation in  Eq.\eqref{a6} by using
NU method. Introducing a new variable $z=\cos\theta$  and
$\gamma=2(E+M)$ and inserting these into Eq.\eqref{a6} yield
\ba \Theta ''(z)-\frac{2z}{1-z^2}\Theta
'(z)+\frac{1}{(1-z^2)^2}\left[\lambda
(1-z^2)-m^2-\gamma(\beta^\prime +\beta z)\right]\Theta (z)=0.
\label{a53} \ea
After the comparison of Eq.\eqref{a53}  with Eq.\eqref{a14} we
obtain
\ba \tilde{\tau}(z)=-2z~, \sigma (z)=1-z^2~~,\tilde{
\sigma}(z)=-\lambda z^2-\gamma \beta z+(\lambda -m^2-\gamma
\beta^\prime ). \label{a54} \ea
In the NU method the new function $\pi (z)$ is calculated for
angle-dependent part as
\ba \pi (z)=\pm \sqrt {z^2(\lambda -k)+\gamma \beta z -(\lambda
-\gamma \beta^\prime -m^2-k)}. \label{a55} \ea
The constant parameter $k$ can be determined as
\ba k_{1,2}=\frac{2\lambda -m^2-\gamma \beta^\prime }{2}\pm
\frac{u}{2}, \label{a56} \ea
where $u=\sqrt {(m^2+\gamma \beta^\prime )^2-\gamma^2\beta ^2}$. The
appropriate function $\pi (z)$ and parameter $k$ are
\ba \pi (x ) =  - \left[ x \sqrt {\frac{m^2  + \gamma \beta^\prime +
u}{2}} + \sqrt {\frac{m^2  + \gamma \beta^\prime  - u}{2}} \right],
\label{a57} \ea
\ba k = \frac{2\lambda - m^2 - \gamma \beta^\prime }{2} -
\frac{u}{2}. \label{a58} \ea
The following track in this selection is to achieve the condition
$\tau'<0$ . Therefore  $\tau(z)$ becomes 
\ba \tau (z) =  - 2z\left[ {1 +\sqrt {\frac{{m^2 + \gamma
\beta^\prime + u}}{2}} } \right] - 2\sqrt {\frac{{m^2  + \gamma
\beta^\prime -u}}{2}}. \label{a59} \ea

We can also write the values  $\bar\lambda =k+\pi'(s)$ as
\ba \bar \lambda =\frac{2\lambda -\gamma
\beta^\prime-m^2}{2}-\frac{u}{2}-\sqrt{\frac{m^2+\gamma \beta^\prime
+u}{2}}~, \label{a60} \ea
and using Eq.\eqref{a31}, then from the Eq.\eqref{a60} we can obtain
\ba \bar \lambda_{N} =\frac{2\lambda -\gamma
\beta^\prime-m^2}{2}-\frac{u}{2}-\sqrt{\frac{m^2+\gamma \beta^\prime
+u}{2}}=2N\left[1+\sqrt{\frac{m^2+\gamma \beta^\prime
+u}{2}}\right]+N(N-1). \label{a61} \ea
In order to obtain unknown $\lambda $ we can solve Eq.\eqref{a61}
explicitly for $\lambda =l(l+1)$
\ba \lambda -\zeta ^2-\zeta =2N(1+\zeta )+N(N-1), \label{a62} \ea
where $\zeta =\sqrt{\frac{m^2+\gamma \beta^\prime +u}{2}},$ and
\ba \lambda =\zeta ^2+\zeta +2N\zeta +N(N+1)=(N+\zeta )(N+\zeta
+1)=l(l+1), \label{a63} \ea
then \ba l=N+\zeta. \label{a64} \ea
Substitution of this result Eq.\eqref{a64} in
Eqs.({\ref{a33}-\ref{a35}})  yields the desired energy spectrum, in
terms of $n_r$ and $l$ quantum numbers. Similarly, the wave function
of polar angle dependent part of KFG equation can be formally
derived by a process to the derivation of radial part of KFG
equation. Thus using Eq.\eqref{a19}, we obtain
\ba \phi (z)=(1-z)^{(B+C)/2} , \label{a65} \ea
where $ B=\sqrt{\frac{m^2+\gamma \beta^\prime
+u}{2}}~,~C=\sqrt{\frac{m^2+\gamma \beta^\prime -u}{2}}~. $

On the other hand, to find a solution for $y_N(s)$ we should first
obtain the weight function $\rho(s)$. From Pearson equation, we find
weight function as
\ba \rho (z)=(1-z)^{B+C}(1+z)^{B-C}~. \label{a66} \ea

Substituting Eq.\eqref{a66} into Eq.\eqref{a39} allows us to obtain
the polynomial $y_N (s)$ as follows
\ba y_{N} (z) = B_N (1 -z )^{-(B + C)}(1 + z )^{C - B} \frac{{d^N
}}{{dz^N }}\left[ {(1 - z)^{B + C + N} (1 +z )^{B - C + N}}
\right]~. \label{a67} \ea

From the definition of Jacobi polynomials, we can write
\ba \frac{d^N}{dz^N}\left[(1-z)^{B+C+N}(1+z)^{B-C+N}\right]=(-1)^N
2^N (1-z)^{B+C}(1+z)^{B-C}P_N^{(B + C,B - C)}(z)~. \label{a68} \ea
Having inserted Eq.\eqref{a68} into  Eq.\eqref{a67} and after long
but straightforward calculations we obtain the following result,
\ba \Theta_{N}(z) = C_N (1 - z)^{(B + C)/2} (1 + z)^{(B - C)/2}
P_N^{(B + C,B - C)}(z), \label{a69} \ea
where $C_N$ is the normalization constant. Using orthogonality
relation of the Jacobi polynomials ~\cite{Abramowitz} the
normalization constant can be found as
\ba C_N=\sqrt{\frac{(2N+2B+1)\Gamma (N+1)\Gamma
(N+2B+1)}{2^{2B+1}\Gamma (N+B+C+1)\Gamma (N+B-C+1)}}~. \label{a70}
\ea
Thus after inserting  Eq.\eqref{a64} in Eqs.({\ref{a33}-\ref{a35}})
then we directly obtain energy  spectrum for combined potential, so
Hulth\'en plus Ring-Shaped potential in this form:

In case $V(r)\neq S(r)$ we find
\ba M^2-E_{n_{r},l}^2 = \left[
\frac{\alpha^2-\eta-1/2-n_r(n_r+1)-(2n_{r}+1)\sqrt{\frac{1}{4}+\beta^2+\eta}
}{2n_r + 1 + 2\sqrt{\frac{1}{4}+\beta^2+\eta}}\cdot\delta \right]^2
- C_0 \eta\delta^2 \label{a100} \ea

\noindent In case $V_0=S_0$, then for energy spectrum we obtain:
\ba M^2-E_{n_{r},l}^2 =
\left[\frac{\alpha^2}{2(n_r+N+\zeta+1)}-\frac{(n_r+N+\zeta+1)}{2}
\right]^2 \cdot\delta^2- C_0\eta\delta^2. \label{a101} \ea

\noindent In this case $\beta^2 =0$, but $\alpha^2
=\frac{2V_0(E+M)}{\delta^2}$.

\noindent In case $V(r)=-S(r)$, then for energy spectrum we obtain:
\ba M^2-E_{n_{r},l}^2 =\left[
\frac{\alpha'^2}{2(n_r+N+\zeta+1)}-\frac{(n_r+N+\zeta+1)}{2}
\right]^2\cdot\delta^2- C_0\eta\delta^2. \label{a102} \ea

\noindent In this case also $\beta^2 =0$, but $\alpha'^2\neq
\alpha^2$, here

\ba \alpha'^2 =\frac{{2V_0(E-M)}}{\delta^2}\label{a37} \ea

In the equations Eqs.({\ref{a100}-\ref{a102}})  we used notation
$\eta=(N+\zeta)(N+\zeta+1) $

\section{\bf The Solution of Klein-Fock-Gordon equation for Hulth\'en potential  within SUSY quantum mechanics}\label{sr}

The eigenfunction of ground state $\chi_{0}(r)$ in Eq.\eqref{a5}
according to supersymmetric quantum mechanics, is a form as
\ba
 \chi_{0}(r)=Nexp\left(-\int
W(r)dr\right). \label{a71} \ea
\noindent where $N$ and $W(r)$ are normalized constant and
superpotential, respectively. The connection between the
supersymmetric partner potentials $V_{1}(r)$ and $V_{2}(r)$ of the
superpotential $W(r)$ is as follows ~\cite {Cooper1,Cooper2}:
\ba V_{1} (r)=W^{2} (r)- W'(r)+E\, \, ,\, \, \, \, V_{2} (r)=W^{2}
(r)+ W'(r)+E . \label{a72} \ea
The particular solution of the Riccati equation Eq.\eqref{a72}
searches the following form:
\ba W(r)=-\left(C+\frac{De^{-\delta r}}{1-e^{-\delta r}}\right),
\label{a73} \ea
\noindent where $C$ and $D$ unknown constants. Since $V_{1}
(r)=V_{\rm eff}(r)$, having inserted the relations Eq.\eqref{a118}
and Eq.\eqref{a73} into the expression Eq.\eqref{a72}, and from
comparison of compatible quantities in the left and right sides of
the equation, we find the following relations for $C$ and $D$
constants:
\ba C^2=\varepsilon^2\delta^2+\delta^{2}C_{0}l(l+1), \label{a74} \ea
\ba 2CD-\delta D=\delta^2 l(l+1)-\alpha^2 \delta^2, \label{a75} \ea
\ba
 D^2-\delta D = \delta^2 l(l+1)+\delta^2 \beta^2.
 \label{a76}
\ea
Considering extremity conditions to wave functions, we obtain $D>0$
and  $C<0$.
\noindent Solving Eq.\eqref{a76} yields
\ba D=\frac{\delta\pm\sqrt{\delta^2+4\delta^2 l(l+1)+\beta^2
\delta^2}}{2}=\frac{\delta\pm
2\delta\sqrt{\frac{1}{4}+l(l+1)+\beta^2}}{2}, \label{a77} \ea
and considering $B>0$ from Eqs.\eqref{a75} and \eqref{a76}, we find
\ba 2CD-D^2=-\alpha^2 \delta^2-\beta^2\delta^2, \label{a78} \ea
or
\ba C=\frac{D}{2}-\frac{(\alpha^2+\beta^2)\delta^2}{2D}, \label{a79}
\ea
From  Eq.\eqref{a74} and  Eq.\eqref{a79}, we find
\ba
\varepsilon^2\delta^2+l(l+1)\delta^2C_0=\frac{1}{\delta^2}\left[\frac{D}{2}-\frac{(\alpha^2+\beta^2)\delta^2}{2D}\right]^2,
\label{a80} \ea
After inserting \eqref{a80} into \eqref{a10} for energy eigenvalue,
we obtain
\ba M^2-E^2=
\left[\frac{D}{2}-\frac{(\alpha^2+\beta^2)\delta^2}{2D}\right]^2-l(l+1)C_0\delta^2
\label{a81} \ea
In Eq. \eqref{a78} insert $D>0$ from Eq.\eqref{a74} finally, for
energy eigenvalue, we obtain
\ba
M^2-E^2=\delta^2\left[\frac{1+2\sqrt{\frac{1}{4}+\beta^2+l(l+1)}}{4}-\frac{\alpha^2+\beta^2}{\sqrt{\frac{1}{4}+\beta^2+l(l+1)}}\right]^2-l(l+1)C_0\delta^2
\label{a82} \ea
When $r\rightarrow\infty$, the chosen superpotential  $W(r)$ is
$W(r)\rightarrow -\frac{\hbar A}{\sqrt{2\mu}}$.

\noindent Having inserted the Eq.\eqref{a73} into Eq.\eqref{a72},
then we can find supersymmetric  partner potentials $V_{1}(r)$ and
$V_{2}(r)$ in the form
\ba
V_{1}(r)= W^2 (r) - W'(r)+E = \nn \\
=\biggl[ C^2  + \frac{(2CD -\delta
D)e^{-\delta r}}{1 - e^{-\delta r}} + \nn \\
\frac{(D^2-\delta D)e^{-2\delta r}}{(1 - e^{-\delta r} )^2}\biggr].
\label{a83} \ea
\ba
V_{2}(r) &=& W^2 (r) + W'(r)+E = \nn \\
&&=\left[ C^2  + \frac{(2CD +\delta D)e^{-\delta r}}{1 - e^{ -
\delta r}} + \frac{(D^2+\delta D)e^{-2\delta r}}{(1 - e^{ - \delta
r} )^2}\right], \label{a84} \ea
By using superpotential $W(r)$ from Eq.\eqref{a71} we can find
$\chi_{0}(r)$ radial eigenfunction in this form: \ba
\chi_{0}(r)=N_{0}exp\left(-\int
W(r)dr\right)=N_{0}exp\left[\int\left(C+\frac{De^{ - \delta
r}}{1-e^{ - \delta r}}\right)dr\right]=\nn \\
N_{0}e^{Cr}exp\left[\frac{D}{\delta}\int\frac{d(1-e^{ - \delta
r})}{1-e^{ - \delta r}}\right]= N_{0}e^{Cr}(1-e^{ - \delta
r})^{\frac{D}{\delta}}, \label{a85} \ea
here $r\rightarrow 0$; $\chi_{0}(r)\rightarrow0, D>0$ and
$r\rightarrow \infty$; $\chi_{0}(r)\rightarrow0, C<0$

Two partner potentials $V_{1}(r)$ and $V_{2}(r)$ which differ from
each other with additive constants and have the same functional form
are called invariant potentials ~\cite{Gendenshtein1,
Gendenshtein2}. Thus, for the partner potentials $V_{1}(r)$ and
$V_{2}(r)$ given with  Eq.\eqref{a83} and Eq.\eqref{a84}, the
invariant forms are:
\ba R(D_1 ) = V_{2}(D,r) - V_{1}(D_1 ,r) = \left[C^2 -C_{1}^2\right] =  \nn \\
=\left[\frac{D}{2}-\frac{(\alpha^2+\beta^2)\delta ^2}{2D}\right]^2 -
\left[\frac{D +
\delta}{2}-\frac{(\alpha^2+\beta^2)\delta^2}{2(D+\delta)} \right]^2.
\label{a86} \ea
\ba
R(D_i ) = V_{2}[D + (i - 1)\delta ,r] - V_{1}[D + i\delta ,r] = \nn \\
\left[
\left(\frac{D+i\delta}{2}-\frac{(\alpha^2+\beta^2)\delta^2}{2(D+i\delta)}\right)^2
- \left(\frac{D +
(i-1)\delta}{2}-\frac{(\alpha^2+\beta^2)}{2(D+(i-1)\delta)}
\right)^2\right]. \,\,\,\,\,\,\,\,\,\,\,\, \label{a87} \ea
where the reminder $R(D_i)$ is independent of $r$.

If we continue this procedure and make the substitution
$\,D_{n_r}=D_{n_r-1} +\delta =D+n_r\delta$ at every step until $\,
D_{n} \ge 0$, the whole discrete spectrum of Hamiltonian
$\,H_{-}(D)$:

\ba
E_{n_{r}l} &=& E_0  + \sum\limits_{i = 0}^n R(D_i ) = \nn \\
&&=\lambda \delta^2 C_0 - \left(\frac{D}{2} -
\frac{\delta^2(\alpha^2+\beta^2)}{2D} \right)^2-
\biggl[\left(\frac{D+\delta}{2}-
\frac{\delta^2(\alpha^2+\beta^2)}{2(D+\delta)} \right)^2-  \nn \\
&&-\left(\frac{D}{2} - \frac{\delta^2(\alpha^2+\beta^2)}{2D}
\right)^2 + \left(\frac{D+2\delta}{2} -
\frac{\delta^2(\alpha^2+\beta^2)}{2(D+2\delta)}\right)^2+.....+ \nn \\
&&+\left(\frac{D+(n_r-1)\delta}{2} -
\frac{\delta^2(\alpha^2+\beta^2)}{2(D+(n_r-1)\delta)}\right)^2 -
\biggl(\frac{D+(n_r-2)\delta}{2} - \nn \\
&& -\frac{\delta^2(\alpha^2+\beta^2)}{2(D+(n_r-2)\delta)}\biggr)^2-
\left(\frac{D+(n_r-2)\delta}{2} - \frac{\delta^2
(\alpha^2+\beta^2)}{2(D+(n_r-2)\delta)}\right)^2 +  \nn \\
&&+\left(\frac{D+n_r\delta}{2}-
\frac{\delta^2(\alpha^2+\beta^2)}{2(D+n_r\delta)}\right)^2- \nn \\
&&-\left(\frac{D+(n_r-1)\delta}{2} -
\frac{\delta^2(\alpha^2+\beta^2)}{2(D+(n_r-1)\delta)}\right)^2\biggr] =  \nn \\
&&=\lambda \delta^2 C_0- \left(\frac{D+n_r\delta}{2} -
\frac{\delta^2(\alpha^2+\beta^2)}{2(D+n_r\delta)} \right)^2,
\label{a88} \ea
and we obtain
\ba
 E_{n_{r}l} &=& \lambda\delta^2 C_0-
\frac{{\hbar^2}}{{2\mu}} \left(\frac{D+n_r\delta}{2} -
\frac{\delta^2(\alpha^2+\beta^2)}{2(D+n_r\delta)} \right)^2,
\label{a89} \ea
Finally, for energy eigenvalues we found
\ba M^2-E_{n_{r},l}^2 = \left[
\frac{\alpha^2-\lambda-1/2-n(n+1)-(2n+1)\sqrt{\frac{1}{4}+\beta^2+\lambda}
}{ 2n_r + 1 + 2\sqrt{\frac{1}{4}+\beta^2+\lambda}}\cdot\delta
\right]^2-\nn \\ l(l+1) C_0 \delta^2. \label{a90} \ea
In case $V_0=S_0$, then for energy spectrum we obtain:
\ba M^2-E_{n_{r},l}^2 = \left[
\frac{\alpha^2-\lambda-1/2-n(n+1)-(2n+1)\sqrt{\frac{1}{4}+\lambda}
}{ 2n_r + 1 + 2\sqrt{\frac{1}{4}+\lambda}}\cdot\delta \right]^2
-l(l+1) C_0 \delta^2= \nn \\
\left[ \frac{\alpha^2}{2(n+l+1)}-\frac{(n+l+1)}{2}
\right]^2\cdot\delta^2 -l(l+1) C_0 \delta^2. \label{a91} \ea

In this case $\beta^2 =0$

In case $V(r)=-S(r)$, then for energy spectrum we obtain:
\ba M^2-E_{n_{r},l}^2 = \left[
\frac{\alpha^2-\lambda-1/2-n(n+1)-(2n+1)\sqrt{\frac{1}{4}+\lambda}
}{ 2n_r + 1 + 2\sqrt{\frac{1}{4}+\lambda}}\cdot\delta \right]^2
-l(l+1) C_0 \delta^2= \nn \\
\left[ \frac{\alpha'^2}{2(n+l+1)}-\frac{(n+l+1)}{2}
\right]^2\cdot\delta^2 -l(l+1) C_0 \delta^2. \label{a92} \ea

In this case also $\beta^2 =0$, but $\alpha'^2\neq \alpha^2$, here

\ba \alpha'^2 =\frac{2V_0(E-M)}{\delta^2}\label{a93} \ea

Based on Eqs.(A.14) and (A .17), the obtained result of radial KFG
equation by using the Eq.\eqref{a71} of the ground state
eigenfunction is exactly same with the result obtained by using NU
method.

This indisputable opens a new window for determining of the
properties of the interactions in quantum system.

\section{\bf Numerical Results and Discussion}\label{nr}

Solutions of the modified  Klein-Fock-Gordon equation for the
Hulth\'en plus ring-shaped like potential are obtained respectively
within ordinary quantum mechanics by applying the Nikiforov-Uvarov
method and within SUSY QM by applying the shape invariance concept.
Both ordinary and SUSY quantum mechanical energy eigenvalues and
corresponding eigenfunctions  are obtained for arbitrary $l$ quantum
numbers.

After analytically  solving the bound states of $l$-wave KFG
equation with vector and scalar Hulth\'en plus ring-shaped
potentials, we should make next  important remarks. First, when $l =
0$, the approximation centrifugal term
${l(l+1)\delta^2(C_0+\frac{e^{-\delta r}}{(1-e^{\delta r})^2})}=0$,
too. Thus letting $l = 0$ in Eq.\eqref{a34} and Eq.\eqref{a49}, they
reduce to the exact energy spectrum formula and the unnormalized
radial wave functions for the bound states of $s$-wave KFG equation
with vector and scalar Hulth\'en potentials
\cite{Adame,Chetouani,Chen}, respectively. Second, in the case
scalar potential is equal to the vector potential, as $S_0=V_0$, and
$\zeta=0$ in Eq.\eqref{a101} then formula for energy spectrum
reduces to Eq.\eqref{a34}
\ba M^2-E_{n_{r},l}^2 =
\left[\frac{\alpha^2}{2(n_r+l+1)}-\frac{(n_r+l+1)}{2} \right]^2
\cdot\delta^2- C_0 l(l+1)\delta^2. \nn  \ea

In this case $\beta^2 =0$.

Third, in case $V(r)=-S(r)$, and $\zeta=0$ in Eq.\eqref{a102} then
for energy spectrum we obtain Eq.\eqref{a36}:
\ba M^2-E_{n_{r},l}^2 =\left[
\frac{\alpha'^2}{2(n_r+l+1)}-\frac{(n_r+l+1)}{2}
\right]^2\cdot\delta^2- C_0 l(l+1)\delta^2.\nn  \ea

\noindent In this case also.

Fourth, we also discussed non-relativistic limit of the  formula for
the energy spectrum. When $V(r)=S(r)$, then Eq.\eqref{a5}  will be
transformed a Schr\"{o}dinger equation for the potential $2V(r)$. If
we take $C_0=0$ in Eq.\eqref{a34}, and also $C_0=0$ and $\zeta=0$ in
Eq.\eqref{a100} then we obtain results \cite{Yuan,Qiang}.

If $\zeta=0$, then Eqs.({\ref{a100}-\ref{a102}}) reduces to energy
spectrum for the Hulth\'en potential.

The energy eigenvalues and corresponding eigenfunctions are obtained
for arbitrary $l$ quantum numbers. Two important cases must be
emphasized in the results of this study. In the first case which $
\beta =\beta ^\prime=0$ the potentials turn to central Hulth\'en
potential. For this case, by using $ u=m^2$, $\zeta =|m|$ and $l=N
+|m|~(N=0,1,2...)$ then $l \geq | m| $ by substituting this $l$
values in Eq.\eqref{a100} we obtain energy spectrum for Hulth\'en
potential.

Finally, we want to deal with some restrictions about bound state
solutions of KFG for Hulth\'en plus ring shaped like potential.
First, it is seen from  Eq.\eqref{a56} and expression from $u$ that
in order to obtain real energy values the condition $(m^2+\beta
')^2\geq \beta ^2 $ must be hold. Since the parameters $ \beta $ and
$\beta '$ are real and positive, we can write \ba
 m^2 \geq \gamma(\beta-\beta') .
\label{a103} \ea

If $\beta \leq \beta'$ the inequality in Eq.\eqref{a103}  is
provided automatically. But if $\beta \geq \beta'$ then $m$ becomes
bounded. Secondly, in  Eq.\eqref{a34} if

\ba l(l+1) C_0>\left[
\frac{\alpha^2-\lambda-1/2-n_r(n_r+1)-(2n_{r}+1)\sqrt{\frac{1}{4}+\beta^2+\lambda}
}{ 2n_r + 1 + 2\sqrt{\frac{1}{4}+\beta^2+\lambda}}\cdot\right]^2
\label{a104} \ea

\noindent then energy eigenvalues  take  non-negative values, this
means there is no bound states.

If both conditions in  Eqs.({\ref{a103}-\ref{a104}})  are satisfied
simultaneously, the bound states exist.

\section{\bf Conclusion}\label{cr}

We  used alternative two methods to obtain the energy eigenvalues
and corresponding eigenfunctions of the Klein-Fock-Gordon equation
for the Hulth\'en plus ring-shaped lake potential.

The energy eigenvalues of the bound states and corresponding
eigenfunctions are analytically found via both of NU and SUSY
quantum mechanics. The same expressions were obtained for the energy
eigenvalues, and the expression of radial and azimuthal wave
functions transformed each other is  shown by using these methods. A
closed form of the normalization constant of the wave functions is
also found. The energy eigenvalues and corresponding eigenfunctions
are obtained for arbitrary $l$  angular momentum and $n_r$  radial
quantum numbers. It is shown that the energy eigenvalues and
eigenfunctions are sensitive to $n_r$ radial and $l$ orbital quantum
numbers.

It is worth to mention that the Hulth\'en plus ring-shaped lake
potential   is one of the important exponential potential, and it is
a subject of interest in many fields of physics and chemistry. The
main results of this paper are the explicit and closed form
expressions for the energy eigenvalues and the normalized wave
functions. The method presented in this study  is a systematic one
and in many cases it is one of the most concrete works in this area.

Consequently, studying of analytical solution of the modified KFG
equation is obtained for the Hulth\'en plus ring-shaped lake
potential within the framework ordinary and SUSY QM could provide
valuable information on the QM dynamics at nuclear, atomic and
molecule physics and opens new window.

We can conclude that our analytical results of this study are
expected to enable new possibilities for pure theoretical and
experimental physicist, because the results are exact and more
general.

\appendix

\section{Supersymmetric Quantum Mechanics}

Supersymmetric Quantum Mechanics (SUSYQM) for $N=2$, we have two
nilpotent operators namely, $Q$ and $Q^{+} $, satisfying the
following algebra:
\ba \{ Q,\, Q^{+} \}=H\, ,\, \, \, \, \, \, \{ Q\, ,\, Q\}=\{ Q^{+}
,Q^{+} \} =0, \ea where $H$ is the
supersymmetric Hamiltonian, $Q=\left(\begin{array}{cc} {0} & {0} \\
{A^{-} } & {0}
\end{array}\right)$ and $Q^{+} =\left(\begin{array}{cc} {0} & {A^{+}
} \\ {0} & {0} \end{array}\right)$ are the operators of
supercharges, $A^{-} $ is bosonic operators and $A^{+}$ is its
adjoint. The supersymmetric $H$ Hamiltonian in terms of these
operators defined in the form  ~\cite{Cooper1,Cooper2}:
\ba H=\left(\begin{array}{cc} {A^{+} A^{-} } & {0} \\ {0} & {A^{-}
A^{+}
} \end{array}\right)\, =\left(\begin{array}{cc} {H_{-} } & {0} \\
{0} & {H_{+} }
\end{array}\right),
\ea
\noindent where $H_{-} $ and $H_{+} $ are called supersymmetric
partner Hamiltonians. The supercharges $Q$ and $Q^{+} $ commute with
SUSY $H$ Hamiltonian: $[H\, ,\, Q]=[H,Q^{+} ]=0$.

If the ground state energy of a Hamiltonian $H$ is zero (i.e.
$E_{0}=0$), it can always be written in a factorable form as a
product of a pair of linear differential operators. That is why, one
has from the Schr\"{o}dinger equation that the ground state wave
function $\psi _{0} (x)$ obeys
\ba H\psi _{o}(x)=-\frac{\hbar^{2} }{2m} \frac{d^{2} \psi _{0}
}{dx^{2}} +V(x)\psi _{0} (x)=0, \ea
so that
\ba V(x)=\frac{\hbar ^{2} }{2m} \frac{\psi ''_{0} (x)}{\psi _{0}
(x)}. \ea
This allows a global reconstruction of the potential $V(x)$ from the
knowledge of its ground state wave function which possesses no
nodes. Once we realize this, factorizing the Hamiltonian is now
quite simple by using the following ansatz ~\cite{Cooper1,Cooper2}:
/%
\ba H_{-}=-\frac{\hbar ^{2} }{2m} \frac{d^{2} }{dx^{2} } +V(x)=A^{+}
A^{-} \ea
 here
\ba A^{-} =\frac{\hbar}{\sqrt{2m} } \frac{d}{dx} +W(x)\,, \,  A^{+}
=-\frac{\hbar }{\sqrt{2m} } \frac{d}{dx} +W(x). \ea
By factorizing procedure of the Hamiltonian, the Riccati equation
for Superpotential  is obtained:
\ba V_{-}(x)=W^{2} (x)-\frac{\hbar }{\sqrt{2m} } W'(x). \ea
The solution for superpotential $W(x)\, $ in terms of the ground
state wave function is
\ba W(x)=-\frac{\hbar ^{2} }{2m} \frac{\psi '_{0} (x)}{\psi _{0}
(x)}. \ea

This solution is obtained by recognizing that once we satisfy $A^{-}
\psi _{0} (x)=0$, we automatically have a solution to $H\psi _{0}
=A^{+} A^{-} \psi _{0} =0 \,.$

The next step in constructing the SUSY theory related to the
original Hamiltonian $H_{-} $ is to define the operator $H_{+}
=A^{-} A^{+} $ obtained by reversing the order of $A^{-} $ and
$A^{+} $. A little simplification shows that the operator $H_{+} $
is in fact a Hamiltonian corresponding to a new potential $V_{+}
(x)$.
\ba H_{+} =-\frac{\hbar ^{2} }{2m} \frac{d^{2} }{dx^{2} } +V_{+}
(x)\, \, \, ,\, \, \, \, V_{+} (x)=W^{2} (x)+\frac{\hbar }{\sqrt{2m}
} W'(x). \ea
The potentials $V_{-} (x)$ and $V_{+} (x)$ are known as
supersymmetric partner potentials. It is then clear that if the
ground state energy of a Hamiltonian $H_{1}$ is $E^{1}_{0}$ with
eigenfunction $\psi _{0}^{1}$ then in view of Eq.(A.5), it can
always be written in the form below as,
\ba H_{1}=-\frac{\hbar ^{2} }{2m} \frac{d^{2} }{dx^{2} } +V_{1}
(x)=A^{+} A^{-} +E_{0}^{1}, \ea
 here
\ba
\begin{array}{l} {A_{1}^{-}
=\frac{\hbar }{\sqrt{2m} } \frac{d}{dx} +W_{1} (x)\, ,\, \,
A_{1}^{+} =-\frac{\hbar }{\sqrt{2m} } \frac{d}{dx} +W_{1} (x),}
\\ V_{1} (x)=W_{1}^{2} (x)-\frac{\hbar }{\sqrt{2m} } W'_{1}
(x)+E_{0}^{1} , W_{1} (x)=-\frac{\hbar^{2}}{2m}\frac {d \ln
\psi_{0}^{1}}{dx} \,.\end{array} \ea
The SUSY partner Hamiltonian is then given by
~\cite{Cooper1,Cooper2}
\ba H_{2} =A_{1}^{-} A_{1}^{+} +E_{0}^{1} =-\frac{\hbar ^{2} }{2m}
\frac{d^{2} }{dx^{2} } +V_{2} (x), \ea
where
\ba V_{2} (x)=W_{1}^{2} (x)+\frac{\hbar }{\sqrt{2m} } W'_{1}
(x)+E_{0}^{1} = \nn \\
 = V_{1} (x)+\frac{2\hbar }{\sqrt{2m} } W'_{1} (x)=
V_{1}(x)-\frac{\hbar }{m} \frac{d^{2} }{dx^{2} } (\ln \psi
_{0}^{(1)}). \ea

From Eq.(A.12), the energy eigenvalues and eigenfunctions of the two
Hamiltonians $H_{1}$ and $H_{2}$ are related by
~\cite{Cooper1,Cooper2}
\ba E_{n}^{2} =E_{n+1}^{1} \, ,\, \, \, \, \psi
_{n}^{2}=[E_{n+1}^{1} -E_{0}^{1} ]^{-\frac{1}{2}}A_{^{1} }^{-} \psi
_{n+1}^{1}, \nn \\
\psi _{n+1}^{1} =[E_{n}^{2} -E_{0}^{2} ]^{-\frac{1}{2}}A_{^{1} }^{+}
\psi _{n}^{2}. \ea
Here $E_{n}^{m}$ is the energy level, where $n$ denotes the energy
level and $(m)$ refers to the $m$'th Hamiltonian $H_{m}$.

Thus, it is clear that if the original Hamiltonian $H_{1}$ has $p\ge
1$ bound states with eigenvalues $E_{n}^{1}$, and eigenfunctions
$\psi _{n}^{1}$ with $0<n<p$, then we can always generate a
hierarchy of $(p-1)$ Hamiltonians $H_{2} \, ,\, \, H_{3} \, ,\,
...,\, H_{p}$ such that the $m$'th member of the hierarchy of
Hamiltonians $(H_{m})$ has the same eigenvalue spectrum as $H_{1}$
except that the first $(m-1)$ eigenvalues of $H$ are missing in $H$
~\cite{Cooper1,Cooper2}:
\ba H_{m} =A_{m}^{+} A_{m}^{-} +{\rm \; E}_{{\rm m-1}}^{{\rm 1}}
=-\frac{\hbar ^{2} }{2m} \frac{d^{2} }{dx^{2} } +V_{m} (x), \ea
where
\ba A_{m}^{-} =\frac{\hbar }{\sqrt{2m} } \frac{d}{dx} +W_{m} (x)\,
,\, \, \, \, W_{m} (x)=-\frac{\hbar }{\sqrt{2m} } \frac{d\ln
\psi _{0}^{(m)} }{dx}, \nn \\
 m=2\,\,3\,\, 4,\,
\, \cdots \, \, p. \ea
One also has
\ba
\begin{array}{l} E_{n}^{(m)}=E_{n+1}^{(m-1)}=\cdots =E_{n+m-1}^{1} \,, \\ \psi _{n}^{(m)} =[E_{n+m-1}^{1} -E_{m-2}^{1}
]^{-\frac{1}{2}} \cdots [E_{n+m-1}^{1} -E_{0}^{1}]^{-\frac{1}{2}}
A_{m-1}^{-} \cdots A_{1}^{-} \psi^{1}_{n+m-1}, \,\,\,\, \\
V_{m}(x)=V_{1}(x)-\frac{\hbar }{m} \frac{d^{2} }{dx^{2} } \ln (\psi
_{0}^{(1)} \cdots \psi _{0}^{(m-1)}).
\end{array}
\ea
i.e., knowing all the eigenvalues and eigenfunctions of $H_{1}$ we
immediately know all the energy eigenvalues $E_{n}^{1}$ and
eigenfunctions $\psi _{n}^{1}$ of the hierarchy of $(p-1)$
Hamiltonians $H_{2} \, ,\, \, H_{3} \, ,\, ...,\, H_{p}.$

\section*{\bf Acknowledgments}

This work was supported by the Science Development Foundation under
the President of the Republic of Azerbaijan -\textbf{Grant No.
EIF/MQM/Elm-Tehsil-1-2016-1(26)-71/11/1} and \textbf{Grant No
EIF-KETPL-2-2015-1(25)-56/02/1}. A. I. Ahmadov also is grateful for
the financial support  Baku State University \textbf{Grant No.
"50+50" (2018 - 2019)}.


\end{document}